\documentclass[12pt,letterpaper]{article}

\usepackage[letterpaper,margin=1in]{geometry}

\usepackage{amssymb,mathtools} % loads \Bbbk
\usepackage{newtxtext,newtxmath} 

\usepackage{amsmath}
\usepackage{setspace}

\usepackage{comment}
\usepackage[numbers]{natbib}
\usepackage{caption}
\usepackage{threeparttable}
\usepackage{tabularx}
\usepackage{booktabs}
\usepackage[dvipsnames]{xcolor}
\usepackage{array}
\usepackage{soul}
\usepackage{subcaption}
\usepackage{tikz}
\usepackage{pgfplots}
\pgfplotsset{compat=1.18}
\usepackage[hidelinks]{hyperref}
\usepackage{url}
 \usepackage{subcaption}
\usepackage{algpseudocode}
\usepackage{algorithm}
\usepackage{xspace}

 \usepackage{algorithm}

\usepackage{threeparttable}  
\usepackage{multirow}

\algdef{SE}[DOWHILE]{Do}{doWhile}{\algorithmicdo}[1]{\algorithmicwhile\ #1}%

 \usepackage{tikz}
 \usetikzlibrary{plotmarks}
 \usepackage{pgfplots}
 \usetikzlibrary{patterns}
 \pgfplotsset{compat=1.3}
 \usetikzlibrary{backgrounds}

\newcommand{\node}{\ensuremath{a}}

 \newcommand{\PC}[1]{
 	\vspace{2px}
 	\noindent{\bf \IfEndWith{#1}{:}{#1}{#1:}}
 }

\usepackage{etoolbox}
\newcommand{\circled}[2][]{%
	\tikz[baseline=(char.base)]{%
		\node[shape = circle, draw, fill=red, color=red, inner sep = .2pt]
		(char) {\phantom{\ifblank{#1}{#2}{#1}}};%
		\node at (char.center) {\makebox[0pt][c]{\color{white}{#2}}};}}
\robustify{\circled}

\newcommand{\lrdp}{\texttt{L}\text{-}\texttt{RDP}}
\newcolumntype{R}{>{\raggedleft\arraybackslash}X}

\usepackage{amsthm}
\newtheorem{definition}{Definition}
\newtheorem{theorem}{Theorem}
\newtheorem{lemma}{Lemma}

\theoremstyle{remark}
\newtheorem{remark}{Remark}
\theoremstyle{definition}

\renewcommand{\qedsymbol}{$\square$}
\providecommand{\Halmos}{\qedsymbol}
\newenvironment{APPENDICES}{\appendix}{}
\usepackage{titlesec}
\titleformat{\section}
  {\bfseries\fontsize{13}{16}\selectfont}{ }{0pt}{}
\titleformat{\subsection}
  {\bfseries\itshape\fontsize{12}{14}\selectfont}{ }{0pt}{}
\titleformat{\subsubsection}
  {\bfseries\fontsize{12}{14}\selectfont}{ }{0pt}{}

\titlespacing*{\section}{0pt}{12pt}{3pt}
\titlespacing*{\subsection}{0pt}{10pt}{2pt}
\titlespacing*{\subsubsection}{0pt}{8pt}{2pt}

\makeatletter
\makeatletter
\renewcommand{\maketitle}{%
  \begin{center}
    {\bfseries\fontsize{20}{20}\selectfont \@title \par}
    \vspace{1em}
    {\normalsize \lineskip .5em
      \begin{tabular}[t]{c}
        \@author
      \end{tabular}\par
    }
    \vspace{1em}
    {\normalsize \@date \par}
  \end{center}
}
\makeatother

\title{Local Differential Privacy for Federated Learning with Fixed Memory Usage and Per-Client Privacy}
\author{%
\begin{minipage}{0.28\linewidth}
\centering
\singlespacing
Rouzbeh Behnia \\
University of South Florida \\
\texttt{behnia@usf.edu}
\end{minipage}
\and
\begin{minipage}{0.25\linewidth}
\centering
\singlespacing
Jeremiah Birrell \\
Texas State University \\
\texttt{jbirrell@txstate.edu}
\end{minipage}
\and
\begin{minipage}{0.2\linewidth}
\centering
\singlespacing
Arman Riasi \\
Virginia Tech \\
\texttt{armanriasi@vt.edu}
\end{minipage}
\and
\begin{minipage}{0.28\linewidth}
\centering
\singlespacing
Reza Ebrahimi \\
University of South Florida \\
\texttt{ebrahimim@usf.edu}
\end{minipage}
\and
\begin{minipage}{0.28\linewidth}
\centering
\singlespacing
Kaushik Dutta \\
University of South Florida \\
\texttt{duttak@usf.edu}
\end{minipage}
\and
\begin{minipage}{0.25\linewidth}
\centering
\singlespacing
Thang Hoang \\
Virginia Tech \\
\texttt{thanghoang@vt.edu}
\end{minipage}
}

\newenvironment{myabstract}{
  \begin{center}
  \begin{minipage}{0.9\linewidth}
  \begin{center}\textbf{Abstract}\end{center}
}{\end{minipage}\end{center}\vspace{1em}}

\AtBeginEnvironment{align}{\setlength{\abovedisplayskip}{3pt}\setlength{\belowdisplayskip}{3pt}}
\AtBeginEnvironment{align*}{\setlength{\abovedisplayskip}{3pt}\setlength{\belowdisplayskip}{3pt}}

\usepackage{fancyhdr}
\begin{document}
%%%%%%%%%%%%%%%%
\date{}
\maketitle

\begin{myabstract}

Federated learning (FL) enables organizations to collaboratively train models without sharing their datasets. Despite this advantage, recent studies show that both client updates and the global model can leak private information, limiting adoption in sensitive domains such as healthcare. Local differential privacy (LDP) offers strong protection by letting each participant privatize updates before transmission. However, existing LDP methods were designed for centralized training and introduce challenges in FL, including high resource demands that can cause client dropouts and the lack of reliable privacy guarantees under asynchronous participation. These issues undermine model generalizability, fairness, and compliance with regulations such as HIPAA and GDPR. To address them, we propose \lrdp{}, a DP method designed for LDP that ensures constant, lower memory usage to reduce dropouts and provides rigorous per-client privacy guarantees by accounting for intermittent participation.

\end{myabstract}
\vspace{-18pt}
\section{Introduction}

Federated learning (FL) enables model training across decentralized data sources without requiring the data to be shared \citep{pmlr-v54-mcmahan17a}. In FL, clients train local models on their private data and send updates to a central server, which aggregates them into a global model over multiple rounds. However, recent attacks have shown that both client updates and resulting global models can leak substantial information about the clients' data \citep{CarliniTWJHLRBS21, carlini2023extracting}, limiting the adoption of FL in privacy-sensitive domains such as healthcare and finance \citep{insurancefraud}. Differential privacy (DP) has been shown to effectively mitigate this risk via a privacy accountant that tracks the cumulative leakage and proportionally injects noise during training~\citep{AbadiCGMMT016}. In FL, DP can be implemented in two ways: centrally (CDP), where the server adds noise during aggregation, or locally (LDP), where each client perturbs its update before transmission \citep{geyer2017differentially, zhou2020differentially}. LDP offers stronger end-to-end privacy, as client updates are never exposed, making it especially suitable for privacy-sensitive domains \citep{geyer2017differentially}.

Existing LDP methods are developed using conventional DP accountants originally designed for centralized settings~(e.g., \citep{zhou2020differentially}). For instance, a widely used FL framework, Flower \citep{FlowerFL}, adopted in financial \citep{flower2024seriesa, schreyer2022federated} and healthcare \citep{flower2025healthcare} domains, utilizes   RDP \citep{RDP-Mironov17, Opacus} as its DP accountant to offer LDP. However, adopting these methods, originally designed for centralized settings, in LDP introduces two key limitations: (1) 
These DP accountants (e.g., RDP \citep{RDP-Mironov17} and PRV \citep{GopiLW21}) rely on Poisson subsampling, which produces variable-size minibatches. As a result, each client samples a random number of data points in each batch, causing a significant fluctuation in memory usage. For example, memory usage can vary between 7 and 21 GB for a dataset size of 20,000 samples with a batch size of 120. This unpredictability can prevent clients with lower computational capacity from completing local training or participating in FL. (2) These accountants are not designed to track client-level privacy in FL, where participation is asynchronous. Without per-client tracking, they may under- or overestimate true privacy loss.

The practical consequences of these limitations are especially acute in sensitive domains, where they can compromise both model generalizability and privacy \citep{zhang2025towards, topaloglu2021pursuit}. Disparities in computational resources and their impact on participation and model quality/generalizability in FL are well-studied \citep{FL_Hardware}. However, the adoption of LDP can further  exacerbated these challenges. For example, medical institutions with lower computational capacity are more likely to drop out or fail to complete training due to the unpredictable computational demands imposed by LDP with variable batch size \citep{zhang2025towards, xu2021federated}.
This underrepresentation can lead to global models that generalize poorly to patients from those institutions, exacerbating disparities in care \citep{zhang2025towards, xu2021federated}. Additionally, the lack of per-client tracking in existing LDP methods results in inaccurate privacy accounting at the client level. Without precise measurement of individual privacy loss, these methods may either expose sensitive information, raising the potential for regulatory violations (e.g., HIPAA  \citep{topaloglu2021pursuit}), or inject excessive noise, needlessly degrading model utility. Both failures undermine trust and limit the practical viability of deploying FL in sensitive domains such as healthcare.

% While privacy is central to trustworthy FL, it is not sufficient on its own. In sensitive domains like healthcare, ensuring model integrity, that is, verifying that the global model was computed honestly and exclusively from client updates, is equally important. This concern is especially salient in the LDP setting, where clients perturb their updates before transmission precisely because the server is untrusted. Yet if the server cannot be trusted with client updates, it also cannot be assumed to perform aggregation faithfully \citep{zhao2025federation}. Prior work shows that compromised servers can manipulate training by omitting valid updates or injecting adversarial ones, leading to biased or even backdoored models \citep{kairouz2021advances, ding2024identifying, huang2024harmful}. In the medical domain, such tampering could degrade diagnostic performance or introduce dangerous behaviors into clinical decision systems. Verifying that the global model reflects only the intended (and privatized) updates is particularly difficult under LDP, where the updates are not directly observable. While full transparency might enable verification, releasing client updates or proprietary models is rarely feasible in regulated or commercial settings \citep{yang2018applied, appleFL}.

Another vulnerability in existing LDP-based FL stems from the fact that clients perturb their updates before transmission precisely because the server is not trusted to handle unprotected updates, yet the server is still assumed to perform aggregation honestly \citep{zhou2020differentially}. Prior studies have demonstrated that compromised servers can manipulate the global model by selectively omitting client updates or injecting adversarial ones, resulting in poisoned or backdoored models \citep{kairouz2021advances, ding2024identifying, huang2024harmful}. For example, in medical applications, such manipulations can compromise diagnostic accuracy or introduce unsafe behavior into clinical systems \citep{zhang2025towards}. A study by \cite{insurancefraud} demonstrates how backdoored models in medical FL systems can be exploited to commit fraud. 
% Verifying the global model is honestly computed from the clients updates is particularly challenging under LDP, where those updates are private. 
A straightforward solution would be to release all client updates for public audit, but this is rarely feasible in practice, particularly in regulated or commercial settings where client updates are private and the global model is proprietary or confidential \citep{yang2018applied, appleFL}.

To address these challenges, we build upon the work in \citep{FS_RDP} to design a fixed-size RDP accountant tailored for FL with LDP. Unlike conventional accountants based on Poisson subsampling, our method, called \lrdp, offers fixed-size minibatches, resulting in stable memory usage. It also provides accurate per-client privacy tracking under asynchronous participation in FL, ensuring rigorous privacy guarantees. We accompany \lrdp~with a novel model verification method based on multi-party computation (MPC)~\citep{rosulek2018}, which is commonly adopted to attain integrity and privacy in financial applications \citep{hastings2023privacy}. Our method enables participating clients to verify the model generated by the server at each iteration. The verification ensures that the model was computed solely from the updates provided by participating clients, thereby preventing the server from tampering with model integrity. Our approach requires clients to share only encrypted versions of their updates with one another, protecting both the privacy of individual updates and the commercial value of the resulting model.

The fixed-size minibatches and per-client privacy tracking offered by \lrdp{} make its privacy bounds more conservative than those of conventional DP accountants (e.g., RDP \citep{RDP-Mironov17} and PRV \citep{GopiLW21}), requiring more noise to achieve the same privacy level. However, our evaluation across multiple datasets and application domains shows that the resulting utility loss is negligible (around 1\%), making this a favorable trade-off in practice. To further support this claim, we conduct a sensitivity analysis that demonstrates the robustness of our method under varying FL parameters, such as the number of users and training iterations. We have integrated our privacy accountant (\lrdp) and model integrity verification method into  Flower~\citep{FlowerFL}, a widely-used FL framework. The full artifact will be open-sourced for public verification and testing upon acceptance of the paper.

Our method contributes to the development of generalizable and high-quality AI models in sensitive domains (e.g., healthcare) by enabling broader participation in heterogeneous environments where participants vary in their access to computing resources, a challenge that is common even in interorganizational FL settings \citep{zhang2025towards}. This helps mitigate disparities in model performance across underrepresented populations. Our solution offers rigorous, per-client privacy guarantees and model integrity verification, directly addressing regulatory and compliance concerns (e.g., HIPAA). As a result, it lowers institutional barriers to collaboration and fosters trustworthy FL ecosystems.

\section{Preliminaries}

\subsection{Differential Privacy (DP)}

DP guarantees that the presence or absence of any single data point has only a limited influence on the algorithm’s output, thereby protecting the privacy of the data. 

\begin{definition}[Differential Privacy \citep{dwork2006our}] A randomized mechanism \( \mathcal{M} \colon \mathbb{D} \to \mathcal{R} \) satisfies \emph{\((\varepsilon,\delta)\)-differential privacy} if, for all adjacent datasets \( D, D' \in \mathbb{D} \), and for all measurable subsets \( \mathcal{S} \subseteq \mathcal{R} \), it holds that
$
\mathbb{P}[\mathcal{M}(D) \in \mathcal{S}] \leq e^{\varepsilon} \cdot \mathbb{P}[\mathcal{M}(D') \in \mathcal{S}] + \delta.
$
\end{definition}

Adjacency is defined as the relation $D \sim D'$  between two datasets that differ in exactly one data point. 
%In our setting, datasets are indexed as \( D = \{D_{j,n}\}_{j \in \mathbb{Z}^+,\, n \in \mathbb{Z}^+} \), and adjacency holds if there exists a unique pair \( (j^*, n^*) \in \mathbb{Z}^+ \times \mathbb{Z}^+ \) such that \( D_{j^*,n^*} \sim D'_{j^*,n^*} \) and \( D_{j,n} = D'_{j,n} \) for all \( (j,n) \neq (j^*,n^*) \).
%Differential privacy guarantees that the presence or absence of a single individual's data cannot be confidently inferred from the output of the mechanism. 
The parameter $ \varepsilon \geq 0$  quantifies the worst-case privacy loss, while $\delta \in [0,1]$ allows for a negligible failure probability.
A privacy accountant \citep[e.g.,][]{RDP-Mironov17} is a mechanism for tracking and bounding the cumulative privacy loss incurred by composing multiple differentially private operations. It enables accurate computation of the overall privacy guarantee for a sequence of randomized mechanisms applied to data.

% \subsection{Differentially Private SGD (DP-SGD)}
% A foundational method for applying differential privacy in machine learning is {Differentially Private Stochastic Gradient Descent (DP-SGD)}, introduced by \cite{AbadiCGMMT016}. DP-SGD employs a privacy accountant, known as the Rényi differential privacy \citep{RDP-Mironov17}, to track the cumulative privacy loss throughout training. It then modifies the traditional SGD algorithm to incorporate differential privacy at each gradient update step. Specifically, for each iteration \( n \), a random minibatch \( B_{j,n} \subset D_{j,n} \) is sampled, and per-example gradients \( f_{j,n}(\theta, d_{j,n,i}) \) are computed and clipped to have bounded norm \( \|f_{j,n}\| \leq C_{j,n} \). These clipped gradients are averaged and perturbed with Gaussian noise \( Z_{j,n} \sim \mathcal{N}(0, C_{j,n}^2 \sigma_{j,n}^2 I / |B|_{j,n}^2) \), resulting in the update  $
% \Delta\Theta^D_{j,n}(\theta) \coloneqq \frac{1}{|B|_{j,n}} \sum_{i \in B_{j,n}} f_{j,n}(\theta, d_{j,n,i}) + Z_{j,n}.
% $ 
% By controlling the sampling probability and noise scale, and applying Rényi differential privacy composition theorems, the cumulative privacy loss across training rounds can be bounded. This makes DP-SGD a principled and practical approach for training models on sensitive data, with quantifiable privacy guarantees.

\looseness-1
Differentially Private Stochastic Gradient Descent (DP-SGD)~\citep{AbadiCGMMT016} is a foundational method for training machine learning models with differential privacy. It modifies standard SGD by clipping per-example gradients \( f_{j,n}(\theta, d_{j,n,i}) \) to a norm bound \( C_{j,n} \), averaging them over a random minibatch \( B_{j,n} \subset D_{j,n} \), and adding Gaussian noise \( Z_{j,n} \sim \mathcal{N}(0, C_{j,n}^2 \sigma_{j,n}^2 I / |B|_{j,n}^2) \). The resulting update is:
$
\Delta\Theta^D_{j,n}(\theta) \coloneqq \frac{1}{|B|_{j,n}} \sum_{i \in B_{j,n}} f_{j,n}(\theta, d_{j,n,i}) + Z_{j,n}.
$
RDP accountant~\citep{RDP-Mironov17} tracks cumulative privacy loss, enabling tight composition bounds across training. This makes DP-SGD both practical and theoretically sound for private  training.

% \subsection{Multi-Party Computation (MPC) }
% Similar to prior work that applies MPC to verify computations in financial systems \citep{hastings2023privacy}, our method employs MPC to ensure the integrity of model updates in federated learning, a fundamentally different setting where the goal is to protect and verify the training process of AI models.
%  MPC protocols are built on the foundation of cryptographic secret sharing \citep{shamirSecret}. Secret sharing allows a data owner to divide a secret into multiple pieces, called {shares}, such that each individual share reveals nothing about the original value. These shares are distributed among participants. One of the simplest forms of secret sharing is {additive secret sharing}. To share a secret value $s \in \mathbb{Z}$ among $n$ participants, the data owner samples $n$ values $r_1, \dots, r_n \in \mathbb{Z}$ uniformly at random, subject to the constraint that $s = \sum_{i=1}^{n} r_i$. Each participant $i$ receives the share $r_i$. The crucial property of this scheme is that any subset of fewer than $n$ shares reveals no information about $s$, while the full set of $n$ shares can be used to reconstruct the secret exactly. 

\subsection{Federated Learning (FL) and Local Differential Privacy (LDP)}
We assume a countable universe of possible clients, indexed by $j\in\mathbb{Z}^+$. Let $t\in\mathbb{Z}_0$ denote the server timestep and  $J_t$ be the set of clients (client indices) available at timestep $t$; we assume $|J_t|$ is finite for all $t$. Given minibatch sizes $m_t\leq |J_t|$, we independently select  uniformly random subsets $M_t\subset J_t$ of size $m_t$.
For each client index $j$, we define $N_{j,t}$ be the number of minibatches that client $j$ has been included in, up to timestep $t$, i.e., 
$
  N_{j,t}\coloneqq|\{s\leq t:j\in M_s\}|\,.
$
Note that these are $\sigma(M_s:s\leq t)$-measurable random variables and $N_{j,t}\leq N_{j,t+1}\leq N_{j,t}+1$ for all $t$.
For $n\in\mathbb{Z}^+$ we let $T_{j,n}$ denote server timestep at which client $j$ has been queried for the $n$'th time, i.e., 
$
    T_{j,n}\coloneqq \inf\{t\in\mathbb{Z}_0:N_{j,t}=n\}=\inf\{t\in\mathbb{Z}_0:N_{j,t}\geq n\}\,.
$
We note that  $\{t:N_{j,t}=n\}$ is a.s. nonempty for all $j$, $n$.  Therefore, we can hereafter restrict to the probability-$1$ set where $T_{j,n}$ is finite (and hence the inf is a min) for all $j,n$.  Therefore $N_{j,T_{j,n}}=n$ and if $s\in\mathbb{Z}_0$ with $s<T_{j,n}(\omega)$ then $N_{j,s}(\omega)<n$.  To see that these are $\sigma(M_s:s\leq t)$-measurable, we can rewrite them  as
$
    T_{j,n}=\sum_{t\in\mathbb{Z}_0} t1_{\{N_{j,t}=n\}\cap_{s<t} \{N_{j,s}<n\}}\,.
$

{\bf Update sent to server from client $j$ at query $n$:}  We assume each client has their own dataset, which can change over time; specifically, we let  $D_{j,n}$ denote the client's data  $j$ when it is queried by the server for an update for $n$'th time. We consider the local DP framework, where each client adds their own noise before returning their update to the server, i.e., given the model $\theta$ provided by the server, client $j$ returns the following (random) update when queried by the sever for the $n$th time:
\begin{align}\label{eq:client_Delta_Theta_def}
\Delta\Theta^D_{j,n}(\theta)\coloneqq  \frac{1}{|B|_{j,n}}\sum_{i\in B_{j,n}} f_{j,n}(\theta,d_{j,n,i})+Z_{j,n}\,,
\end{align}
where $d_{j,n,i}$ denotes the $i$'th data point in  $D_{j,n}$ and $B_{j,n}$ is a  uniformly random subset of $D_{j,n}$ having (fixed) size $|B|_{j,n}$, and $Z_{j,n}\sim N(0,C_{j,n}^2\sigma^2_{j,n}I/|B|_{j,n}^2)$. We assume $f_{j,n}$ are measurable in $\theta$ and  satisfy the per-sample non-random clipping bound $\|f_{j,n}\|\leq C_{j,n}$ (as in  clipped SGD). 

{\bf Update at server:} Given an initial model $\Theta_0$, which we will assume does not depend on the clients' data, $D$, the time-$t$ server update is given by
$
\Theta_{t+1}\coloneqq\Theta_t+\frac{1}{m_t}\sum_{j\in M_t} \Delta\Theta_{j,N_{j,t}}(\Theta_t)\,.
$
The random variables $\Theta_0$, $M_t$, $B_{j,n}$, and $Z_{j,n}$ introduced above are all assumed to be independent.   If we want to emphasize the role of $D$ we will write $\Delta\Theta_{j,n}^D$ and $\Theta_t^D$.

\section{Our Proposed LDP Accountant (\lrdp)}\label{sec:LRDP}

In this section, we present our results, which provide the privacy bounds for our accountant, \lrdp, designed for local differential privacy in federated learning. \lrdp\ supports fixed-size subsampling and enables accurate tracking of privacy loss across asynchronous client participation in FL.  
% \subsubsection{\lrdp~under Modification of one Data Point}
Here we analyze the privacy of the mechanism consisting of $N$ updates submitted to the server by a particular client, $j^\prime$, defined by
\begin{align}\label{eq:mechanism_def}
\mathcal{M}^{FS}_{j^\prime,N}(D)\coloneqq\left(\Delta\Theta_{j^\prime,1}^D(\Theta^D_{T_{j^\prime,1}}),...,\Delta\Theta_{j^\prime,N}^D(\Theta^D_{T_{j^\prime,N}})\right)\,.
\end{align}
We will consider an adjacent collection of datasets,  $D^\prime$, where for each $n$, $D^\prime_{j^\prime,n}$ differs from $D_{j^\prime,n}$ by modifying one data point (via add, remove, or replace), and $D^\prime_{j,n}=D_{j,n}$ for all $j\neq j^\prime$ and all $n$. We begin the analysis with the following intuitively clear result, which states that the models held by the server  do not depend on client $j^\prime$'s information, other than through the  sequence of updates it reports to the server.
\begin{lemma}\label{lemma:Hn}
For all $j^\prime\in J$, $n\in\mathbb{Z}^+$,  there is a measurable map $H_{j^\prime,n}$ such that  
\begin{align}\label{eq:Hn_def}
\Theta^D_{T_{j^\prime,n}}=&H_{j^\prime,n}\left(\{D_{j,m}\}_{j\neq j^\prime,m\in\mathbb{Z}_+},W_{\setminus j^\prime},    \{\Delta\Theta_{j^\prime,\ell}^D(\Theta_{T_{j^\prime,\ell}}^D)\}_{\ell=1,...,n-1}\right)\,,\\
W_{\setminus j^\prime}\coloneqq &\left(\Theta_0,\{M_t\}_{t\in\mathbb{Z}_0}, \{(B_{j,m},Z_{j,m})\}_{j\neq j^\prime,m\in\mathbb{Z}^+}\right)\,,\notag
\end{align}
for all $D$. Moreover,
for all $j^\prime$, $n$, the random variables $(B_{j^\prime,n},Z_{j^\prime,n})$ and
    \begin{align}
        \left(W_{\setminus j^\prime},\Delta\Theta_{j^\prime,1}^D(\Theta_{T_{j^\prime,1}}),...,\Delta\Theta_{j^\prime,n-1}^D(\Theta_{T_{j^\prime,n-1}})\right)
        \end{align}
    are independent.
\end{lemma}
\begin{remark}
    We emphasize that the key feature of the $H_{j^\prime,n}$'s is that they do not directly depend on  client $j^\prime$'s random variables,   $\{(B_{j^\prime,m},Z_{j^\prime,m})\}_{m\in\mathbb{Z}^+}$, or data, $\{D_{j^\prime,m}\}_{m\in\mathbb{Z}^+}$; they only depend on these indirectly  through the dependence on $\{\Delta\Theta_{j^\prime,\ell}^D(\Theta_{T_{j^\prime,\ell}}^D)\}_{\ell=1,...,n-1}$. The random quantity $W_{\setminus j^\prime}$ consists of all the random variables used (and controlled) by either the server or a client other than $j^\prime$, and is thus independent from client $j^\prime$'s random variables, $\{(B_{j^\prime,m},Z_{j^\prime,m})\}_{m\in\mathbb{Z}^+}$.
\end{remark}
\begin{proof}{Proof.}
We will prove  that for $s\leq t$ there exists a map $H_{j^\prime,n,t,s}$  such that
\begin{align}\label{eq:H_nts}
    \Theta_s^D1_{T_{j^\prime,n}=t}
    =&H_{j^\prime,n,t,s}\left(\{D_{j,m}\}_{j\neq j^\prime,m\in\mathbb{Z}_+},W_{\setminus j^\prime},    \{\Delta\Theta_{j^\prime,\ell}^D(\Theta_{T_{j^\prime,\ell}}^D)\}_{\ell=1,...,n-1})\right)\,.
\end{align}
Once this is shown, then the first claimed result follows from the equality
\begin{align}
        \Theta^D_{T_{j^\prime,n}}=\Theta_01_{T_{j^\prime,n}=0}+\sum_{t\in\mathbb{Z}^+} \Theta^D_t 1_{T_{j^\prime,n}=t}\,.
    \end{align}

Now, for each $t$, we show \eqref{eq:H_nts} by induction on $s$.  For $s=0$ it follows from the fact that the $T_{j,n}$'s are functions of $\{M_t\}_{t\in \mathbb{Z}_0}$. Now suppose it holds for some $s<t$. Noting that $\{T_{j^\prime,n}=t\}\cap\{N_{j^\prime,s}=\ell\}=\emptyset$ when $s<t$ and $\ell\geq n$, we can compute
    \begin{align}
&\Theta^D_{s+1} 1_{T_{j^\prime,n}=t}\\
=&\Theta^D_{s} 1_{T_{j^\prime,n}=t}+\frac{1}{m_s}\sum_{j\in M_s} 1_{T_{j^\prime,n}=t} \Delta \Theta^D_{j,N_{j,s}}(\Theta^D_s1_{T_{j^\prime,n}=t})\notag\\
=&\Theta^D_{s} 1_{T_{j^\prime,n}=t}+\frac{1}{m_s}\sum_{j\in M_s,j\neq j^\prime} 1_{T_{j^\prime,n}=t} \Delta \Theta^D_{j,N_{j,s}}(\Theta^D_s1_{T_{j^\prime,n}=t})\notag\\
&+\frac{1}{m_s}1_{j^\prime\in M_s} 1_{T_{j^\prime,n}=t} \sum_{\ell<n}1_{N_{j^\prime,s}=\ell}\Delta \Theta^D_{j^\prime,\ell}(\Theta^D_s)\notag\\
=& H_{j^\prime,n,t,s}+\frac{1}{m_s}\sum_{j\in M_s,j\neq j^\prime} 1_{T_{j^\prime,n}=t} \sum_{\ell} 1_{N_{j,s}=\ell}\left(\frac{1}{|B|_{j,\ell}}\sum_{i\in B_{j,\ell}} f_{j,\ell}(H_{j^\prime,n,t,s},d_{j,\ell,i})+Z_{j,\ell}\right)
\notag\\
&+\frac{1}{m_s}1_{j^\prime\in M_s} 1_{T_{j^\prime,n}=t} \sum_{\ell<n}1_{N_{j^\prime,s}=\ell}\Delta \Theta^D_{j^\prime,\ell}(\Theta^D_{T_{j^\prime,\ell}})\notag\\
\coloneqq & H_{j^\prime,n,t,s+1}\left(\{D_{j,m}\}_{j\neq j^\prime,m\in\mathbb{Z}_+},W_{\setminus j^\prime},    \{\Delta\Theta_{j^\prime,\ell}^D(\Theta_{T_{j^\prime,\ell}}^D)\}_{\ell=1,...,n-1}\right)\,,\notag
    \end{align}
where we used the fact that $T_{j^\prime,\ell}=s$ on  $\{j^\prime\in M_s\}\cap\{N_{j^\prime,s}=\ell\}$ and that the $N_{j,t}$'s are functions of $\{M_t\}_{t\in\mathbb{Z}_0}$. This proves the first claim by induction.

To conclude the independence claim, note that the above computation, together with the definition \eqref{eq:client_Delta_Theta_def} implies that
\begin{align}\left(W_{\setminus j^\prime},\Delta\Theta_{j^\prime,1}^D(\Theta_{T_{j^\prime,1}}),...,\Delta\Theta_{j^\prime,n-1}^D(\Theta_{T_{j^\prime,n-1}})\right)
\end{align}
is a measurable function of   $(W_{\setminus j^\prime},\{(B_{j^\prime,\ell},Z_{j^\prime,\ell})\}_{\ell<n})$, which is independent  from $(B_{j^\prime,n}, Z_{j^\prime,n})$. This completes the proof. \Halmos
\end{proof}

Using the Lemma \ref{lemma:Hn} we can decompose the distribution of the mechanism  $\mathcal{M}^{FS}_{j^\prime,N}(D)$ from \eqref{eq:mechanism_def} into the composition of the distribution of $(W_{\setminus j^\prime},\mathcal{M}^{FS}_{j^\prime,N-1}(D))$ with the probability kernel
\begin{align}
&p^D_{j^\prime,N}(\Delta \theta_N|w,\Delta\theta_1,...,\Delta\theta_{N-1})\\
\sim&  \frac{1}{|B|_{j^\prime,N}}\sum_{i\in B_{j^\prime,N}} f_{j^\prime,N}\left(H_{j^\prime,N}\left(\{D_{j,m}\}_{j\neq j^\prime,m\in\mathbb{Z}_+},w,\Delta\theta_1,...,\Delta\theta_{N-1}\right),d_{j^\prime,N,i}\right)+Z_{j^\prime,N}\,.\notag
\end{align}
We emphasize that for the composition formula to hold at the level of distributions, it is key that we have independence of $(W_{\setminus j^\prime},\mathcal{M}^{FS}_{j^\prime,N-1}(D))$  and  $(B_{j^\prime,N},Z_{j^\prime,N})$, as implied by  Lemma \ref{lemma:Hn}.

In light of the above composition formula, we can follow the strategy from   Theorem 2.1 in \citep{AbadiCGMMT016} and bound the R{\'e}nyi divergence between the mechanisms operating on the adjacent datasets $D$ and $D^\prime$  by taking worst-case bounds over the input state at each update step. This yields
\begin{align}\label{eq:worst_case_Renyi_bound}
  &D_\alpha\left(\mathcal{M}^{FS}_{j^\prime,N}(D)\|\mathcal{M}^{FS}_{j^\prime,N}(D^\prime)\right)\\
  \leq&\sum_{n=1}^N \sup_{w,\Delta\theta_1,...,\Delta\theta_{n-1}} D_\alpha\left(p^D_{j^\prime,n}(\Delta \theta_n|w,\Delta\theta_1,...,\Delta\theta_{n-1})\|p^{D^\prime}_{j^\prime,n}(\Delta \theta_n|w,\Delta\theta_1,...,\Delta\theta_{n-1})\right)\notag\\
  \leq&\sum_{n=1}^N \sup_{\theta} D_\alpha\left(p^{D_{j^\prime,n}}_{j^\prime,n}(\Delta \theta_n|\theta)\|p^{D^\prime_{j^\prime,n}}_{j^\prime,n}(\Delta\theta_{n}|\theta)\right)\,,\label{eq:multistep_bound}\\
  &p^{D_{j^\prime,n}}_{j^\prime,n}(\Delta \theta_n|\theta)\coloneqq\binom{|D_{j^\prime,n}|}{|B|_{j^\prime,n}}^{-1}\sum_{b}N_{\mu_{j^\prime,n}(\theta,b,D_{j^\prime,n}),\Sigma_{j^\prime,n}}(\Delta\theta_n)\,,\label{eq:sum_b}\\
  &\mu_{j^\prime,n}(\theta,b,D_{j^\prime,n})\coloneqq \frac{1}{|B|_{j^\prime,n}}\sum_{i\in b} f_{j^\prime,n}(\theta,d_{j^\prime,n,i})\,,\,\,\,\Sigma_{j^\prime,n}\coloneqq C_{j^\prime,n}^2\sigma^2_{j^\prime,n}I/|B|_{j^\prime,n}^2\,,
  \end{align}
  where the sum over $b$ in \eqref{eq:sum_b} runs over all allowed minibatches (i.e., subsets of sample indices of size $|B|_{j^\prime,n}$).   Note that to obtain the second inequality we used  the fact that  $D^\prime_{j,m}=D_{j,m}$ for all $m$ and all $j\neq j^\prime$, and hence
  \begin{align}
     H_{j^\prime,n}\left(\{D_{j,m}\}_{j\neq j^\prime,m\in\mathbb{Z}_+},w,\Delta\theta_1,...,\Delta\theta_{n-1}\right)
      =&H_{j^\prime,n}\left(\{D^\prime_{j,m}\}_{j\neq j^\prime,m\in\mathbb{Z}_+},w,\Delta\theta_1,...,\Delta\theta_{n-1}\right)\,.
  \end{align}

\begin{remark}
We emphasize that the bound \eqref{eq:worst_case_Renyi_bound} does not rely on the server operating as intended (i.e., honestly).  The same R{\'e}nyi divergence bound holds for the client $j^\prime$ updates, no matter what functions, $H_{j^\prime,n}$, the server uses to send a sequence of models to client $j^\prime$ (i.e., not only for the $H_{j^\prime,n}$'s obtained in Lemma \ref{lemma:Hn}). Therefore the privacy guarantees we derive below will hold no matter how the server operates, as long as the server does not have direct access to private information from client $j^\prime$, i.e., assuming that the $H_{j^\prime,n}$'s depend only on $j^\prime$'s previously reported updates, $\{\Delta\Theta_{j^\prime,\ell}^D(\Theta_{T_{j^\prime,\ell}}^D)\}_{\ell=1,...,n-1}$, along with (any) random quantities that are independent from  $\{(B_{j^\prime,m},Z_{j^\prime,m})\}_{m\in\mathbb{Z}^+}$ and any data other than $\{D_{j^\prime,m}\}_{m\in\mathbb{Z}_+}$. 
\end{remark}

By adapting the proof of Theorem 3.1 in \citep{FS_RDP}, we can bound each term in the sum \eqref{eq:multistep_bound} (i.e., the one-step R{\'e}nyi divergences)  and thus arrive at the following RDP bound under add/remove adjacency.
\begin{theorem}
Given a client $j^\prime$ and adjacent  collections of datasets $D$ and $D^\prime$     (i.e., $\{D^\prime_{j,m}\}_{m\in\mathbb{Z}^+}=\{D_{j,m}\}_{m\in\mathbb{Z}^+}$ for all $j\neq j^\prime$ and, for all $n$, $D_{j^\prime,n}$ and $D^\prime_{j^\prime,n}$ are adjacent datasets under the add/remove relation), define $q_{j^\prime,n}\coloneqq |B|_{j^\prime,n}/|D_{j^\prime,n}|$. We have
    \begin{align}
      &D_\alpha\left(\mathcal{M}^{FS}_{j^\prime,N}(D)\|\mathcal{M}^{FS}_{j^\prime,N}(D^\prime)\right)\\
  \leq&\sum_{n=1}^N  D_\alpha( q_{j^\prime,n} N_{1,\sigma^2_{j^\prime,n}/4}+(1- q_{j^\prime,n})N_{0,\sigma^2_{j^\prime,n}/4}\|N_{0,\sigma^2_{j^\prime,n}/4})\,.\label{eq:Renyi_ub} 
  \end{align}
\end{theorem}
The R{\'e}nyi divergences in \eqref{eq:Renyi_ub} can then be upper bounded using the asymptotically tight result from Appendix B in \citep{FS_RDP}:
\begin{theorem}\label{thm:Renyi_bound}
For any $m\in\mathbb{Z}^+$, $m\geq 3$, $q\in(0,1)$, $\sigma>0$ we have
    \begin{align}\label{eq:Renyi_bound_Taylor}
&D_\alpha( q N_{1,\sigma^2/4}+(1- q)N_{0,\sigma^2/4}\|N_{0,\sigma^2/4})\\
\leq&\frac{1}{\alpha-1}\log\left[1+\sum_{k=2}^{m-1}\frac{q^k}{k!} \left(\prod_{j=0}^{k-1}(\alpha-j)\right) M_{\sigma,k}+R_{\alpha,\sigma,m}(q)\right]\,,\notag\\
&M_{\sigma,k}\coloneqq\sum_{\ell=2}^k(-1)^{k-\ell}\binom{k}{\ell}e^{2\ell(\ell-1 )/\sigma^2}+(-1)^{k-1}(k-1)\,,\label{eq:M_def}
  \end{align}
  where a computable bound on the remainder term $R_{\alpha,\sigma,m}(q)$ can be found in Appendix \ref{app:Renyi_bound}.
\end{theorem}

\section{Evaluations}
\begin{figure*}[t]
  \centering
  \begin{subfigure}{\columnwidth} % or 0.9\columnwidth
    \centering
    \input{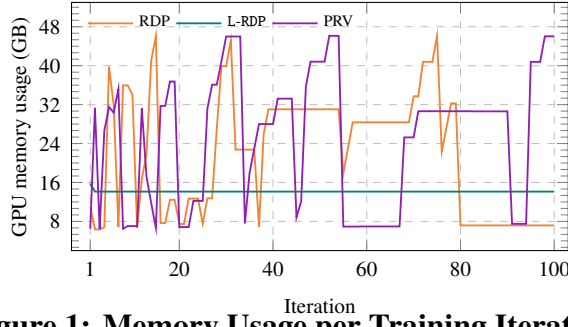}
  \end{subfigure}
  \vspace{-2.5em}
  \caption{Memory Usage per Training Iteration}
  %\vspace{-1.5em}
  \label{fig:expMem}
\end{figure*}
We have implemented our LDP privacy accountant and incorporated it into the widely used FL framework, Flower \citep{FlowerFL}. To evaluate performance, we compare our method (\lrdp) against two widely-adopted DP accountants from Meta’s Opacus library, Renyi Differential Privacy (RDP) \citep{RDP-Mironov17} and Privacy Random Variable (PRV) \citep{GopiLW21}, both of which are also adopted in Flower. We design two experiments. The first demonstrates the constant memory usage of our method compared to existing counterparts in the FL setting. The second evaluates the utility of models privately trained with \lrdp{} against those trained with RDP and PRV.
\subsection{Memory Usage}
 \looseness -2
  In our first experiment, designed to highlight the ramifications of variable mini-batch size caused by Poisson subsampling, we measure the memory usage of \lrdp{} with that of PRV and RDP. Using CIFAR-10 partitioned across clients, with each client holding a local dataset of 30,000 samples and a batch size of 128, the results (Figure~\ref{fig:expMem}) show that \lrdp{} consistently maintains a fixed and substantially lower memory footprint (14 GB), whereas PRV and RDP can reach up to 46 GB. Such high memory usage can significantly increase client dropout rates, particularly among participants with less capable computational resources, thereby undermining model generalizability and fairness.
\subsection{Model Utility}
Our method’s rigorous per-client privacy guarantees, combined with fixed-size subsampling, result in more conservative privacy bounds that necessitate injecting greater noise during training than existing approaches. In our experiments, this translates to roughly twice the noise magnitude compared to conventional DP accountants. Nevertheless, our results show that even with higher noise, the convergence and performance of models trained with \lrdp{} are only negligibly affected. To illustrate this, we report model accuracy across all experiments, comparing our method against its counterparts under varying privacy budgets ($\epsilon = 2, 4, 6, 8, 10$) in two widely used domains: image recognition and natural language understanding (NLU). For all experiments, we select hyperparameters that maximize the accuracy of the corresponding non-private baseline model.
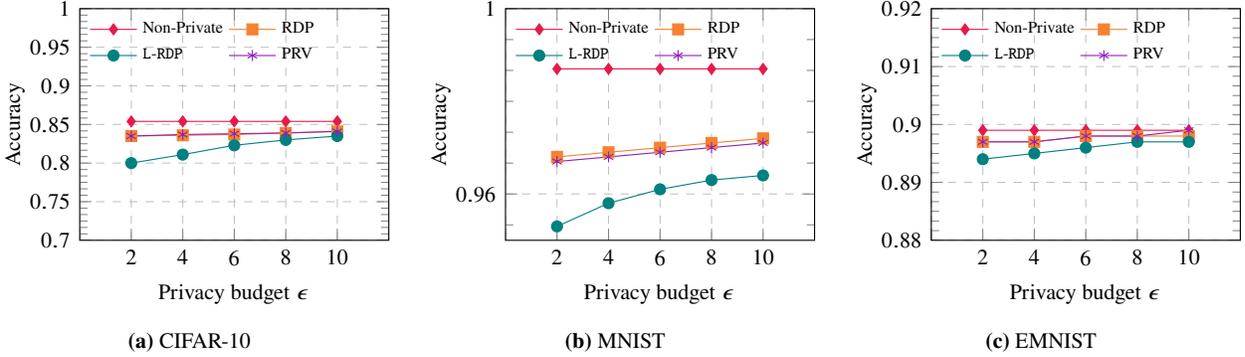
\begin{figure*}[!t] 
	\centering
		\resizebox{1.01\textwidth}{!}{
			\begin{subfigure}{0.3\textwidth}
				% This file was created by matlab2tikz.
%
%The latest updates can be retrieved from
%  http://www.mathworks.com/matlabcentral/fileexchange/22022-matlab2tikz-matlab2tikz
%where you can also make suggestions and rate matlab2tikz.
%
\definecolor{A}{HTML}{e6194B}%
\definecolor{B}{HTML}{f58231}%
\definecolor{C}{HTML}{4363d8}%
\definecolor{D}{HTML}{911eb4}%
\definecolor{E}{HTML}{3cb44b}%
\definecolor{F}{rgb}{0.92900,0.69400,0.12500}%
\definecolor{G}{HTML}{808000}%
\definecolor{H}{HTML}{000000}%
\begin{tikzpicture}
	\footnotesize
	\begin{axis}[%
		width=0.8\textwidth,
		height=0.6\textwidth,
		at={(1.128in,0.894in)},
		scale only axis,
		xmin=0,
		xmax=6,
		xlabel={Privacy budget $\epsilon$},
		xtick={0, 1, 2, 3, 4, 5, 6},
		xticklabels={$ $, 2, 4, 6, 8, 10},
		ymin=0.70,
		ymax=1.0,
            %ymode=log,
		ytick distance=0.05,
		ylabel = {{Accuracy}},
		ylabel shift=-5pt,
		yticklabel shift={0cm},
		axis background/.style={fill=white},
		legend columns=2,
		legend style={legend cell align=left, align=left, fill=none, draw=none,inner sep=-0pt, row sep=0pt, font = \tiny},
		legend pos = north west,
		ymajorgrids,
		xmajorgrids,
            %ymode=log,
		grid style={line width=.5pt, draw=gray!90,dashed},
		major grid style={line width=.2pt,draw=gray!50},
		minor y tick num=5,
            label style={font=\scriptsize},  % Set font size for x and y axis labels
            tick label style={font=\scriptsize}  % Set font size for tick labels      
		]

		\addplot [color=A, solid, mark=diamond*, mark options={solid, A}]
		table[row sep=crcr]{%
			1 0.854 \\ %0.8540
			2 0.854 \\ %0.8540
			3 0.854\\ %0.8540
			4 0.854\\ %0.8540
			5 0.854\\ %0.8540
		};
		\addlegendentry{Non-Private}

		\addplot [color=B, solid, mark=square*, mark options={solid, B}]
		table[row sep=crcr]{%
			1 0.835\\ % 0.8451
			2 0.836\\ % 0.8462
			3 0.837\\ % 0.8470
			4 0.839\\ % 0.8490
			5 0.841\\ % 0.8502
		};
		\addlegendentry{RDP}

		\addplot [color=teal, solid, mark=otimes*, mark options={solid, teal}]
		table[row sep=crcr]{%
			1 0.80\\ %0.7998
			2 0.811\\ %0.8218
			3 0.823\\ %0.8336
			4 0.830\\ %0.8404
			5 0.835\\ %0.8456
		};
		\addlegendentry{\lrdp}

		\addplot [color=D, solid, mark=asterisk, mark options={solid, D}]
		table[row sep=crcr]{%
			1 0.835\\ % 0.8456
			2 0.837\\ % 0.8471
			3 0.838\\ % 0.8483
			4 0.839\\ % 0.8491
			5 0.841\\ % 0.8510
		};
		\addlegendentry{PRV}

    %\draw[solid, black, latex-latex, line width=0.01pt] (515, 11.15) -- (515,14.269);
    %\node[solid, black, latex-latex] at (545, 12.67) {\scriptsize 22x};

    %\draw[solid, black, latex-latex, line width=0.01pt] (515, 6.9) -- (515,11.15);
    %\node[solid, black, latex-latex] at (545, 9.025) {\scriptsize 70x};

	\end{axis}

\end{tikzpicture}%
                    \caption{CIFAR-10}
                    %\label{fig:exp:cifar} 
			\end{subfigure}\hspace{4mm}
			\begin{subfigure}{0.3\textwidth}
				% This file was created by matlab2tikz.
%
%The latest updates can be retrieved from
%  http://www.mathworks.com/matlabcentral/fileexchange/22022-matlab2tikz-matlab2tikz
%where you can also make suggestions and rate matlab2tikz.
%
\definecolor{A}{HTML}{e6194B}%
\definecolor{B}{HTML}{f58231}%
\definecolor{C}{HTML}{4363d8}%
\definecolor{D}{HTML}{911eb4}%
\definecolor{E}{HTML}{3cb44b}%
\definecolor{F}{rgb}{0.92900,0.69400,0.12500}%
\definecolor{G}{HTML}{808000}%
\definecolor{H}{HTML}{000000}%
\begin{tikzpicture}
	\footnotesize
	\begin{axis}[%
		width=0.8\textwidth,
		height=0.6\textwidth,
		at={(1.128in,0.894in)},
		scale only axis,
		xmin=0,
		xmax=6,
		xlabel={Privacy budget $\epsilon$},
		xtick={0, 1, 2, 3, 4, 5, 6},
		xticklabels={$ $, 2, 4, 6, 8, 10},
		ymin=0.95,
		ymax=1.0,
            %ymode=log,
		ytick distance=0.04,
		ylabel = {Accuracy},
		ylabel shift=-5pt,
		yticklabel shift={0cm},
		axis background/.style={fill=white},
		legend columns=2,
		legend style={legend cell align=left, align=left, fill=none, draw=none,inner sep=-0pt, row sep=0pt, font = \tiny},
		legend pos = north west,
		ymajorgrids,
		xmajorgrids,
            %ymode=log,
		grid style={line width=.5pt, draw=gray!90,dashed},
		major grid style={line width=.2pt,draw=gray!50},
		minor y tick num=5,
            label style={font=\scriptsize},  % Set font size for x and y axis labels
            tick label style={font=\scriptsize}  % Set font size for tick labels   
		]

		\addplot [color=A, solid, mark=diamond*, mark options={solid, A}]
		table[row sep=crcr]{%
			1 0.987 \\ %
			2 0.987 \\ %
			3 0.987\\ %
			4 0.987\\ %
			5 0.987\\ %
		};
		\addlegendentry{Non-Private}

		\addplot [color=B, solid, mark=square*, mark options={solid, B}]
		table[row sep=crcr]{%
			1 0.968\\ % 0.9917
			2 0.969\\ % 0.9916
			3 0.970\\ % 0.9919
			4 0.971\\ % 0.9910
			5 0.972\\ % 0.9915
		};
		\addlegendentry{RDP}

		\addplot [color=teal, solid, mark=otimes*, mark options={solid, teal}]
		table[row sep=crcr]{%
			1 0.953\\ % 0.9901
			2 0.958\\ % 0.9909
			3 0.961\\ % 0.9906
			4 0.963\\ % 0.9903
			5 0.964\\ % 0.9905
		};
		\addlegendentry{\lrdp}

		\addplot [color=D, solid, mark=asterisk, mark options={solid, D}]
		table[row sep=crcr]{%
			1 0.967\\ % 0.9916
			2 0.968\\ % 0.9918
			3 0.969\\ % 0.9919
			4 0.970\\ % 0.9915
			5 0.971\\ % 0.9919
		};
		\addlegendentry{PRV}

	\end{axis}

\end{tikzpicture}%
                    \caption{MNIST}
                    %\label{fig:exp:mnist} 
			\end{subfigure}\hspace{4mm}
            \begin{subfigure}{0.3\textwidth}
				% This file was created by matlab2tikz.
%
%The latest updates can be retrieved from
%  http://www.mathworks.com/matlabcentral/fileexchange/22022-matlab2tikz-matlab2tikz
%where you can also make suggestions and rate matlab2tikz.
%
\definecolor{A}{HTML}{e6194B}%
\definecolor{B}{HTML}{f58231}%
\definecolor{C}{HTML}{4363d8}%
\definecolor{D}{HTML}{911eb4}%
\definecolor{E}{HTML}{3cb44b}%
\definecolor{F}{rgb}{0.92900,0.69400,0.12500}%
\definecolor{G}{HTML}{808000}%
\definecolor{H}{HTML}{000000}%
\begin{tikzpicture}
	\footnotesize
	\begin{axis}[%
		width=0.8\textwidth,
		height=0.6\textwidth,
		at={(1.128in,0.894in)},
		scale only axis,
		xmin=0,
		xmax=6,
		xlabel={Privacy budget $\epsilon$},
		xtick={0, 1, 2, 3, 4, 5, 6},
		xticklabels={$ $, 2, 4, 6, 8, 10},
		ymin=0.88,
		ymax=0.92,
            %ymode=log,
		ytick distance=0.01,
		ylabel = {Accuracy},
		ylabel shift=-5pt,
		yticklabel shift={0cm},
		axis background/.style={fill=white},
		legend columns=2,
		legend style={legend cell align=left, align=left, fill=none, draw=none,inner sep=-0pt, row sep=0pt, font = \tiny},
		legend pos = north west,
		ymajorgrids,
		xmajorgrids,
            %ymode=log,
		grid style={line width=.5pt, draw=gray!90,dashed},
		major grid style={line width=.2pt,draw=gray!50},
		minor y tick num=5,
            label style={font=\scriptsize},  % Set font size for x and y axis labels
            tick label style={font=\scriptsize}  % Set font size for tick labels   
		]

		\addplot [color=A, solid, mark=diamond*, mark options={solid, A}]
		table[row sep=crcr]{%
			1 0.899\\ % 0.8991
			2 0.899\\ % 0.8991
			3 0.899\\ % 0.8991
			4 0.899\\ % 0.8991
			5 0.899\\ % 0.8991
		};
		\addlegendentry{Non-Private}

		\addplot [color=B, solid, mark=square*, mark options={solid, B}]
		table[row sep=crcr]{%
			1 0.897\\ % 0.8971
			2 0.897\\ % 0.8975
			3 0.898\\ % 0.8980
			4 0.898\\ % 0.8983
			5 0.898\\ % 0.8982
		};
		\addlegendentry{RDP}

		\addplot [color=teal, solid, mark=otimes*, mark options={solid, teal}]
		table[row sep=crcr]{%
			1 0.894\\ %0.8946
			2 0.895\\ % 0.8957
			3 0.896\\ % 0.8962
			4 0.897\\ % 0.8973
			5 0.897\\ % 0.8978
		};
		\addlegendentry{\lrdp}

		\addplot [color=D, solid, mark=asterisk, mark options={solid, D}]
		table[row sep=crcr]{%
			1 0.897\\ % 0.8973
			2 0.897\\ % 0.8976
			3 0.898\\ % 0.8982
			4 0.898\\ % 0.8987
			5 0.899\\ % 0.8990
		};
		\addlegendentry{PRV}

    %\draw[solid, black, latex-latex, line width=0.01pt] (515, 8.905) -- (515,12.05);
    %\node[solid, black, latex-latex] at (545, 10.4775) {\scriptsize 20x};

    %\draw[solid, black, latex-latex, line width=0.01pt] (515, 5.9) -- (515,8.905);
    %\node[solid, black, latex-latex] at (545, 7.4025) {\scriptsize 22x};

	\end{axis}

\end{tikzpicture}%
                    \caption{EMNIST}
                    %\label{fig:exp:emnist}  
			\end{subfigure}
		}\vspace{-.2em}
  
	\caption{Accuracy vs. Privacy Budget ($\epsilon$) for Image Recognition}\label{fig:exp:IMS} 	
\end{figure*}

For image benchmarks, we evaluate CIFAR-10, MNIST, and EMNIST using a VGG11 architecture with 10 clients. The experimental configurations are as follows: CIFAR-10 with 50 server rounds, five local epochs, and a batch size of 128; MNIST with 25 server rounds, five local epochs, and a batch size of 256; and EMNIST with 25 server rounds, two local epochs, and a batch size of 1024. Across all three image benchmarks, we observe a consistent pattern: the accuracy of \lrdp{} improves steadily with increasing privacy budgets, approaching that of both RDP and PRV despite the more conservative noise injected during the training phase. On CIFAR-10, \lrdp{} achieves 82.3\% accuracy at $\epsilon = 6$ and 83.5\% at $\epsilon = 10$, closely trailing RDP and PRV (both at 84.1\%). On MNIST, the accuracy of \lrdp{} increases from 95.3\% at $\epsilon = 2$ to 96.4\% at $\epsilon = 10$, compared to $\sim$96.8--97.2\% for RDP and PRV over the same range, resulting in a maximum gap of only 1.5\%. On EMNIST, which includes a larger label space and more heterogeneous input patterns, \lrdp{} achieves 89.4\% at $\epsilon = 2$ and 89.7\% at $\epsilon = 10$, closely tracking PRV and RDP (both at 89.7--89.9\%) and effectively matching the non-private model’s accuracy of 89.9\%.

For NLU benchmarks, we evaluate QQP, QNLI, and SST2 using the pre-trained BERT-base  model with 10 clients, five server rounds, and three local epochs. Across all three tasks, we observe that \lrdp{} maintains strong utility. On QQP, \lrdp{} achieves 84.8\% accuracy at $\epsilon = 6$ and 85.8\% at $\epsilon = 10$, compared to 86.8\% for RDP and 86.8–86.9\% for PRV. On QNLI, \lrdp{} improves from 82.9\% at $\epsilon = 2$ to 85.4\% at $\epsilon = 10$, trailing PRV and RDP by at most 1.2 percentage points. On SST2, which exhibits high baseline performance, \lrdp{} reaches 87.7\% at $\epsilon = 8$ and 88.4\% at $\epsilon = 10$, compared to 89.4–89.7\% for RDP and PRV, respectively.

These results demonstrate that although \lrdp{} provides rigorous per-client privacy guarantees, its impact on model utility remains minimal in practice. Notably, the performance gap narrows as the privacy budget increases, suggesting that the relative effect of the additional noise introduced by our method diminishes under less stringent privacy requirements. Overall, \lrdp{} offers a strong balance between privacy and performance, providing rigorous per-client guarantees without compromising the utility of the model.

\begin{figure*}[!t] 
	\centering
		\resizebox{1.01\textwidth}{!}{
			\begin{subfigure}{0.3\textwidth}
				% This file was created by matlab2tikz.
%
%The latest updates can be retrieved from
%  http://www.mathworks.com/matlabcentral/fileexchange/22022-matlab2tikz-matlab2tikz
%where you can also make suggestions and rate matlab2tikz.
%
\definecolor{A}{HTML}{e6194B}%
\definecolor{B}{HTML}{f58231}%
\definecolor{C}{HTML}{4363d8}%
\definecolor{D}{HTML}{911eb4}%
\definecolor{E}{HTML}{3cb44b}%
\definecolor{F}{rgb}{0.92900,0.69400,0.12500}%
\definecolor{G}{HTML}{808000}%
\definecolor{H}{HTML}{000000}%
\begin{tikzpicture}
	\footnotesize
	\begin{axis}[%
		width=0.8\textwidth,
		height=0.6\textwidth,
		at={(1.128in,0.894in)},
		scale only axis,
		xmin=0,
		xmax=6,
		xlabel={Privacy budget $\epsilon$},
		xtick={0, 1, 2, 3, 4, 5, 6},
		xticklabels={$ $, 2, 4, 6, 8, 10},
		ymin=0.7,
		ymax=1.0,
            %ymode=log,
		ytick distance=0.05,
		ylabel = {{Accuracy}},
		ylabel shift=-5pt,
		yticklabel shift={0cm},
		axis background/.style={fill=white},
		legend columns=2,
		legend style={legend cell align=left, align=left, fill=none, draw=none,inner sep=-0pt, row sep=0pt, font = \tiny},
		legend pos = north west,
		ymajorgrids,
		xmajorgrids,
            %ymode=log,
		grid style={line width=.5pt, draw=gray!90,dashed},
		major grid style={line width=.2pt,draw=gray!50},
		minor y tick num=5,
            label style={font=\scriptsize},  % Set font size for x and y axis labels
            tick label style={font=\scriptsize}  % Set font size for tick labels      
		]

		\addplot [color=A, solid, mark=diamond*, mark options={solid, A}]
		table[row sep=crcr]{%
			1 0.8694 \\ %4.43 min
			2 0.8694 \\ %6.84
			3 0.8694 \\ %13.90
			4 0.8694 \\ %14.08
			5 0.8694 \\ %16.70
		};
		\addlegendentry{Non-Private}
  
		\addplot [color=B, solid, mark=square*, mark options={solid, B}]
		table[row sep=crcr]{%
			1 0.8630\\ % 
			2 0.8599 \\ % 
			3 0.8630\\ % 
			4 0.8676\\ % 
			5 0.8682\\ % 
		};
		\addlegendentry{RDP}

		\addplot [color=teal, solid, mark=otimes*, mark options={solid, teal}]
		table[row sep=crcr]{%
			1 0.8342 \\ %4.43 min
			2 0.8478 \\ %6.84
			3 0.8486 \\ %13.90
			4 0.8539 \\ %14.08
			5 0.8584 \\ %16.70
		};
		\addlegendentry{\lrdp}

            \addplot [color=D, solid, mark=asterisk, mark options={solid, D}]
		table[row sep=crcr]{%
			1 0.8642\\ % 
			2 0.8664\\ % 
			3 0.8678\\ % 
			4 0.8680\\ % 
			5 0.8684\\ % 
		};
		\addlegendentry{PRV}

    %\draw[solid, black, latex-latex, line width=0.01pt] (515, 11.15) -- (515,14.269);
    %\node[solid, black, latex-latex] at (545, 12.67) {\scriptsize 22x};

    %\draw[solid, black, latex-latex, line width=0.01pt] (515, 6.9) -- (515,11.15);
    %\node[solid, black, latex-latex] at (545, 9.025) {\scriptsize 70x};

 	\end{axis}

\end{tikzpicture}%
                    \caption{QQP}
                    %\label{fig:exp:cifar} 
			\end{subfigure}\hspace{4mm}
			\begin{subfigure}{0.3\textwidth}
				% This file was created by matlab2tikz.
%
%The latest updates can be retrieved from
%  http://www.mathworks.com/matlabcentral/fileexchange/22022-matlab2tikz-matlab2tikz
%where you can also make suggestions and rate matlab2tikz.
%
\definecolor{A}{HTML}{e6194B}%
\definecolor{B}{HTML}{f58231}%
\definecolor{C}{HTML}{4363d8}%
\definecolor{D}{HTML}{911eb4}%
\definecolor{E}{HTML}{3cb44b}%
\definecolor{F}{rgb}{0.92900,0.69400,0.12500}%
\definecolor{G}{HTML}{808000}%
\definecolor{H}{HTML}{000000}%
\begin{tikzpicture}
	\footnotesize
	\begin{axis}[%
		width=0.8\textwidth,
		height=0.6\textwidth,
		at={(1.128in,0.894in)},
		scale only axis,
		xmin=0,
		xmax=6,
		xlabel={Privacy budget $\epsilon$},
		xtick={0, 1, 2, 3, 4, 5, 6},
		xticklabels={$ $, 2, 4, 6, 8, 10},
		ymin=0.76,
		ymax=0.96,
            %ymode=log,
		ytick distance=0.05,
		ylabel = {Accuracy},
		ylabel shift=-5pt,
		yticklabel shift={0cm},
		axis background/.style={fill=white},
		legend columns=2,
		legend style={legend cell align=left, align=left, fill=none, draw=none,inner sep=-0pt, row sep=0pt, font = \tiny},
		legend pos = north west,
		ymajorgrids,
		xmajorgrids,
            %ymode=log,
		grid style={line width=.5pt, draw=gray!90,dashed},
		major grid style={line width=.2pt,draw=gray!50},
		minor y tick num=5,
            label style={font=\scriptsize},  % Set font size for x and y axis labels
            tick label style={font=\scriptsize}  % Set font size for tick labels   
		]

		\addplot [color=A, solid, mark=diamond*, mark options={solid, A}]
		table[row sep=crcr]{%
			1 0.8691 \\ %4.43 min
			2 0.8691 \\ %6.84
			3 0.8691 \\ %13.90
			4 0.8691 \\ %14.08
			5 0.8691 \\ %16.70
		};
		\addlegendentry{Non-Private}

		\addplot [color=B, solid, mark=square*, mark options={solid, B}]
		table[row sep=crcr]{%
			1 0.8304\\ % 
			2 0.8616\\ % 
			3 0.8638 \\ % 
			4 0.8654\\ % 
			5 0.8671\\ % 
		};
		\addlegendentry{RDP}

		\addplot [color=teal, solid, mark=otimes*, mark options={solid, teal}]
		table[row sep=crcr]{%
			1 0.8290 \\ %4.43 min
			2 0.8458 \\ %6.84
			3 0.8482 \\ %13.90
			4 0.8512 \\ %14.08
			5 0.8541\\ %16.70
		};
		\addlegendentry{\lrdp}

		\addplot [color=D, solid, mark=asterisk, mark options={solid, D}]
		table[row sep=crcr]{%
			1 0.8606 \\ % 
			2 0.8658 \\ % 
			3 0.8665 \\ % 
			4 0.8630 \\ % 
			5 0.8683\\ % 
		};
		\addlegendentry{PRV}

	\end{axis}

\end{tikzpicture}%
                    \caption{QNLI}
                    %\label{fig:exp:mnist} 
			\end{subfigure}\hspace{4mm}
            \begin{subfigure}{0.3\textwidth}
				% This file was created by matlab2tikz.
%
%The latest updates can be retrieved from
%  http://www.mathworks.com/matlabcentral/fileexchange/22022-matlab2tikz-matlab2tikz
%where you can also make suggestions and rate matlab2tikz.
%
\definecolor{A}{HTML}{e6194B}%
\definecolor{B}{HTML}{f58231}%
\definecolor{C}{HTML}{4363d8}%
\definecolor{D}{HTML}{911eb4}%
\definecolor{E}{HTML}{3cb44b}%
\definecolor{F}{rgb}{0.92900,0.69400,0.12500}%
\definecolor{G}{HTML}{808000}%
\definecolor{H}{HTML}{000000}%
\begin{tikzpicture}
	\footnotesize
	\begin{axis}[%
		width=0.8\textwidth,
		height=0.6\textwidth,
		at={(1.128in,0.894in)},
		scale only axis,
		xmin=0,
		xmax=6,
		xlabel={Privacy budget $\epsilon$},
		xtick={0, 1, 2, 3, 4, 5, 6},
		xticklabels={$ $, 2, 4, 6, 8, 10},
		ymin=0.77,
		ymax=1.0,
            %ymode=log,
		ytick distance=0.05,
		ylabel = {Accuracy},
		ylabel shift=-5pt,
		yticklabel shift={0cm},
		axis background/.style={fill=white},
		legend columns=2,
		legend style={legend cell align=left, align=left, fill=none, draw=none,inner sep=-0pt, row sep=0pt, font = \tiny},
		legend pos = north west,
		ymajorgrids,
		xmajorgrids,
            %ymode=log,
		grid style={line width=.5pt, draw=gray!90,dashed},
		major grid style={line width=.2pt,draw=gray!50},
		minor y tick num=5,
            label style={font=\scriptsize},  % Set font size for x and y axis labels
            tick label style={font=\scriptsize}  % Set font size for tick labels   
		]

		\addplot [color=A, solid, mark=diamond*, mark options={solid, A}]
		table[row sep=crcr]{%
			1 0.8990 \\ %4.43 min
			2 0.8990 \\ %6.84
			3 0.8990 \\ %13.90
			4 0.8990 \\ %14.08
			5 0.8990 \\ %16.70
		};
		\addlegendentry{Non-Private}
  
		\addplot [color=B, solid, mark=square*, mark options={solid, B}]
		table[row sep=crcr]{%
			1 0.8853 \\ % 
			2 0.8887 \\ % 
			3 0.8910 \\ %  
			4 0.8944 \\ %  
			5 0.8967 \\ % 
		};
		\addlegendentry{RDP}

            \addplot [color=teal, solid, mark=otimes*, mark options={solid, teal}]
		table[row sep=crcr]{%
			1 0.8577 \\ %4.43 min
			2 0.8692\\ %6.84
			3 0.8738\\ %13.90
			4 0.8777\\ %14.08
			5 0.8841\\ %16.70
		};
		\addlegendentry{\lrdp}

		\addplot [color=D, solid, mark=asterisk, mark options={solid, D}]
		table[row sep=crcr]{%
			1 0.8876 \\ % 
			2 0.8922 \\ % 
			3 0.8944 \\ % 
			4 0.8945 \\ % 
			5 0.8969 \\ % 
		};
		\addlegendentry{PRV}

    %\draw[solid, black, latex-latex, line width=0.01pt] (515, 8.905) -- (515,12.05);
    %\node[solid, black, latex-latex] at (545, 10.4775) {\scriptsize 20x};

    %\draw[solid, black, latex-latex, line width=0.01pt] (515, 5.9) -- (515,8.905);
    %\node[solid, black, latex-latex] at (545, 7.4025) {\scriptsize 22x};

qq	\end{axis}

\end{tikzpicture}%
                    \caption{SST2}
                    %\label{fig:exp:emnist}  
			\end{subfigure}
		}\vspace{-.2em}
  
	\caption{Accuracy vs. Privacy Budget ($\epsilon$) for Natural Language Understanding}\label{fig:exp:IMS} 
\end{figure*}
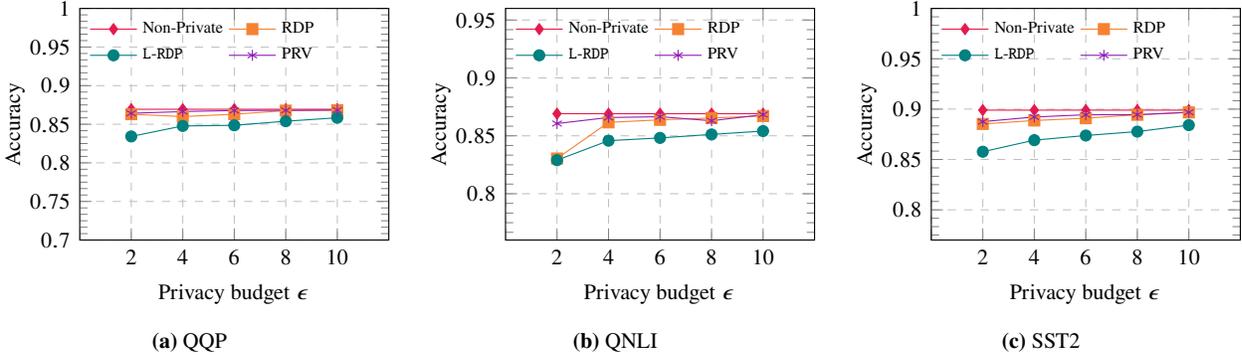

\section{Conclusion and Managerial Implications}
We introduce \lrdp{}, a fixed-size Renyi differential privacy accountant designed specifically for federated learning under the local DP model. Unlike conventional approaches based on Poisson subsampling  \citep[e.g.,][]{RDP-Mironov17}, \lrdp{} ensures significantly lower computational overhead and provides rigorous per-client privacy tracking, even under asynchronous participation. Our experiments across multiple domains show that \lrdp{} addresses the random batch size issue, which directly contributes to user dropouts in existing DP methods, while maintaining high model utility, thereby addressing a key barrier to adoption in privacy-sensitive environments such as healthcare and finance \citep{yan2024federated,FL_Hardware,zhang2025towards}.

% The results carry three important managerial implications. First, by ensuring stable resource usage and minimizing the risk of client dropouts, which is particularly important in settings with high device heterogeneity, \lrdp{} supports more consistent participation in federated learning. This consistency improves the representativeness of training data, which in turn enhances model generalizability and fairness. Second, the ability to track privacy at the level of individual participants provides explicit guarantees that align with regulatory requirements (e.g., HIPAA, GDPR), helping organizations better manage compliance and reduce legal and reputational risks. Third, the balance achieved between negligible utility loss and stronger privacy protection offers decision-makers a practical path for deploying federated learning in heterogeneous environments, lowering institutional barriers to collaboration and fostering trust among stakeholders.

%%%%%%%%%%%%%%%%%%%%%%%%%%%%%%%%%%%%%%%%%%%%%%%%%%%%%%%%%%%%%%%%%%%%%%
% --- References (WITS asks for MISQ reference format; swap style when you have the .bst) ---
{
\singlespacing
\bibliographystyle{IEEEtran}
\bibliography{PPML2025}
% \begin{thebibliography}{}
}
\appendix
\begin{APPENDICES}
\section{R{\'e}nyi Divergence Upper Bound}\label{app:Renyi_bound}
In this appendix, we provide a computable bound on the remainder term, $R_{\alpha,\sigma,m}(q)$, that is needed in order to use the R{\'e}nyi divergence bound from Theorem \ref{thm:Renyi_bound}.  This result is adapted from \citep{FS_RDP}.  We include it here for completeness.

For any $\alpha>1$, $\sigma>0$, write the  R{\'e}nyi divergence on the right-hand side of \eqref{eq:Renyi_bound_Taylor} as
\begin{equation}
 D_\alpha( q N_{1,\sigma_t^2/4}+(1- q)N_{0,\sigma_t^2/4}\|N_{0,\sigma_t^2/4}) =\frac{1}{\alpha-1}\log[H_{\alpha,\sigma_t}(q)]\,,
\end{equation}
where  
\begin{equation}
H_{\alpha,\sigma}(q)\label{H_def}
\coloneqq\!\int \!\left(\frac{ qN_{ 1,\sigma^2/4}(\theta) +(1- q) N_{0,\sigma^2/4}(\theta)}{N_{0,\sigma^2/4}(\theta)}\right)^\alpha\!\!\! N_{0,\sigma^2/4}(\theta)d\theta\,.
\end{equation}
Applying Taylor's formula to $H_{\alpha,\sigma}(q)$ we obtain
\begin{align}\label{eq:H_Taylor_main}
H_{\alpha,\sigma}(q)=\sum_{k=0}^{m-1} \frac{q^k}{k!}H^{(k)}_{\alpha,\sigma}(0)+R_{\alpha,\sigma,m}(q)\,,
\end{align}
where $H^{(k)}_{\alpha,\sigma}$ denotes the $k$'th derivative of $H_{\alpha,\sigma}(q)$ with respect to $q$ and the remainder term in Taylor's formula at order $m$ is given by
\begin{align}\label{eq:remainder_term_main}
R_{\alpha,\sigma,m}(q)\coloneqq 
  q^{m} \int_0^1 \frac{(1-s)^{m-1}}{(m-1)!}H^{(m)}_{\alpha,\sigma}(sq)ds\,.
\end{align}

The following computable bounds on the remainder were derived in \citep{FS_RDP}:
\begin{align}
      &|R_{\alpha,\sigma,m}(q)|\leq \begin{cases}
            q^{m} \prod_{j=0}^{m-1}|\alpha-j| \left[\sum_{\ell=0}^{\lceil\alpha\rceil-m}q^\ell \frac{(\lceil\alpha\rceil-m)!}{(\lceil\alpha\rceil-m-\ell)!(m+\ell)!}   \widetilde{B}_{\sigma,\ell+m}+ \frac{1}{m!}\widetilde{B}_{\sigma,m}\right] &\text{ if }\alpha-m>0\\
\frac{q^{m}}{m!}(1-q)^{\alpha-m}\prod_{j=0}^{m-1}|\alpha-j| \widetilde{B}_{\sigma,m}  &\text{ if }\alpha-m\leq 0\\
  \end{cases}\,.
\end{align}
They are expressed in terms of the  quantities
\begin{align}
  \widetilde{B}_{\sigma,j}\coloneqq  \begin{cases}
         M_{\sigma,j} &\text{ if } j \text{ even}\\
         M_{\sigma,j-1}^{1/2}M_{\sigma,j+1}^{1/2} &\text{ if }j \text{ odd}
     \end{cases}\,,     
\end{align}
where the $M_{\sigma,k}$'s were defined in \eqref{eq:M_def}.

\end{APPENDICES}

\end{document}